\documentclass[conference]{IEEEtran}
	\usepackage{graphicx}
\ifCLASSINFOpdf

\else

\fi
\usepackage{multirow}
\usepackage{mbenotes}

\usepackage{amsmath}
\usepackage{eufrak}
\begin{document}
%
\title{Prototyping and Experimentation of a Closed-Loop Wireless Power Transmission with Channel Acquisition and Waveform Optimization}

\author{\IEEEauthorblockN{Junghoon Kim$^1$, Bruno Clerckx$^1$, and Paul D. Mitcheson$^2$}
\IEEEauthorblockA{$^1$Comm. and Sig. Proc. Group, EEE Department, Imperial College London, United Kingdom\\
$^2$Control and Power Group, EEE Department, Imperial College London, United Kingdom\\
Email:\{junghoon.kim15, b.clerckx, paul.mitcheson\}@imperial.ac.uk}
}


%


\maketitle

\begin{abstract}
A systematic design of adaptive waveform for Wireless Power Transfer (WPT) has recently been proposed and shown through simulations to lead to significant performance benefits compared to traditional non-adaptive and heuristic waveforms. In this study, we design the first prototype of a closed-loop wireless power transfer system with adaptive waveform optimization based on Channel State Information acquisition. The prototype consists of three important blocks, namely the channel estimator, the waveform optimizer, and the energy harvester. Software Defined Radio (SDR) prototyping tools are used to implement a wireless power transmitter and a channel estimator, and a voltage doubler rectenna is designed to work as an energy harvester. A channel adaptive waveform with 8 sinewaves is shown through experiments to improve the average harvested DC power at the rectenna output by 9.8\% to 36.8\% over a non-adaptive design with the same number of sinewaves.\footnote{This work has been partially supported by the EPSRC of UK, under grant EP/P003885/1}
\end{abstract}


%
\IEEEpeerreviewmaketitle

\section{Introduction}
 Recently, interest in energy harvesting has grown rapidly because of the increasing demand for portable electronic devices and the Internet of Things (IoT) technology. RF Wireless Power Transfer (WPT) is becoming a spotlighted area of research in the electrical engineering field. A crucial challenge of WPT designers is to enhance the end-to-end power transfer efficiency, or equivalently enhance the harvested DC power subject to a transmit power constraint. To that end, a major focus in the RF literature has been on enhancing the design of the rectenna so as to increase the RF-to-DC conversion efficiency [1]-[2]. In addition, the transmit signal design has a major impact on the end-to-end power transfer efficiency as it influences the signal strength at the input of the rectenna but also the RF-to-DC conversion efficiency of the rectifier. Some researchers proved that higher rectenna output power could be obtained by using multisine waveforms that exhibit high PAPR [3]-[5]. Unfortunately, those approaches were heuristic and ignored the presence of the wireless propagation channel that is subject to multipath and fading. Moreover, they are based on an open-loop approach with waveform being static. Unlike a wired transmission, in WPT, the input signal of the rectenna at the receiver is not the same as the transmitted signal at the transmitter. The difference between a transmitted signal and a received signal is attributed to (time-varying) reflection, diffraction, and scattering between the transmitter and the receiver. Because of this, the transmitted wireless signal reaches the receiver with multiple different paths of varying lengths, and multipath fading occurs. If the transmission signal is designed without any consideration of multipath fading, the expected power cannot be gathered at the receiver. Therefore, waveform optimization in accordance with channel status is needed to maximize received power and WPT efficiency. A systematic design of channel-adaptive wireless power waveform was proposed in [6]-[7]. This has lead to a closed-loop architecture where the channel state information is acquired to the transmitter and the transmit waveform is dynamically adapted so as to maximize the DC power at the output of the rectifier. Contrary to what is claimed in [3]-[5], maximizing PAPR was shown in [6] not to be a right approach to design efficient wireless power multisine signals. High PAPR multisine signals are good strategies for WPT if the channel is frequency flat, but not in the presence of multipath and frequency selectivity (as in Non Line-of-Sight (NLoS) conditions).  Waveform optimization in [6]-[7] leads to non-convex optimization problems that do not lend themselves easily to implementation. Suboptimal but low complexity waveforms are therefore needed for practical implementation. In [8], a very low complexity channel adaptive waveform design is proposed and shown through simulations to lead to performance very close to the optimal design for a wide range of rectifier topologies.
  
In this paper, we develop and experiment the first prototype of a closed-loop WPT architecture based on channel adaptive waveform optimization and dynamic channel acquisition. We conduct experiments in real-world over-the-air conditions and highlight the benefits of such closed-loop and adaptive architecture over the traditional open-loop non-adaptive design.

\textit{Organization}: Section II introduces the Background theories and section III presents an overall system architecture. Section IV provides experimental measurement results and section V concludes the work and discusses the future plan.

\textit{Notations}: Bold letters stand for vectors or matrices whereas a symbol not in bold font represents a scalar. $\left|.\right|$ and $\left\| . \right\|$ refer to the absolute value of a scalar and the 2-norm of a vector. $\mathcal{E}\{ .\}$ refers to the averaging operator.

\section{Background Theory}
As mentioned in section 1, [6] and [7] introduced a method of multisine waveform optimization relying on Channel State Information (CSI). The researchers of [7] assumed that the transmitter already acquired the CSI perfectly; they then solved the optimization problem by reversed geometric programming (GP) to find amplitudes and phases of multisine waves that maximize the output DC power at the energy harvester. The performance of WPT with the optimized waveform adaptive to the CSI was proved through circuit simulations to have significant gains over various baselines. For example, the simulation results in [6] show that the output power of using channel adaptive and nonadaptive waveforms are 9.5 $\mu$W and 3.5 $\mu$W respectively at -12 dBm input power, 8-tone, and 1 transmit antenna case for a 20MHz bandwidth. At -20dBm with a 10MHz bandwidth, the output power were respectively 2.8 $\mu$W and 1.3 $\mu$W [7]. This shows gains of about 100\% to 200\% of closed-loop approaches over open-loop approaches. Further gains have been observed for a larger number of sinewaves and for multiple transmit antennas [6,7]. But, implementing an optimal waveform design method on a practical system is crucially difficult, because of a long processing time to solve the non-convex optimization problem. If the processing time is too slow, the optimal results would be useless because the channel would have already changed. Therefore, a very low-complexity multisine waveform design strategy called scaled matched filter (SMF) has been proposed in [8] and is considered for implementation in our prototype. A brief introduction to the WPT system model and the SMF method is following.

Consider a multisine transmission signal $x(t)$ written as
\begin{equation}
 x(t) = \mathfrak{R}\{ \sum_{n=0}^{N-1}\omega_{n}e^{j2\pi f_{n}t}\}
\end{equation}
$$\omega_{n} = s_{n}e^{j\phi_{n}}$$
 where $s_{n}$ and $\phi_{n}$ refer to the amplitude and phase of the $n^{th}$ sinewave at frequeny $f_{n}$. Quantities $\mathbf{s}$ and $\Phi$ are N-dimensional vectors of the amplitudes and phases of the sinewaves with their $n^{th}$ entry denoted as $s_{n}$ and $\phi_{n}$. The average transmit power constraint is given by $\mathcal{E}\{ \left|x\right|^{2}\}=\frac{1}{2}\left\| \mathbf{s} \right\|^{2} \le P$. 
 
 The transmitted sinewaves propagate through a multipath channel and the received signal is written as 
\begin{eqnarray}
 y(t) 	&=& \sum_{n=0}^{N-1}s_{n}A_{n}\cos\left(2\pi f_{n}t+\psi_{n} \right) \\
		&=& \mathfrak{R}\{ \sum_{n=0}^{N-1}h_{n}\omega_{n}e^{j2\pi f_{n}t}\}
\end{eqnarray}
 where the channel frequency response at frequency $f_{n}$ is given by
\begin{eqnarray}
h_{n} = A_{n}e^{j\bar{\psi}_{n}}
\end{eqnarray}
with $A_{n}$ and $\psi_{n}$ the amplitude and phase of the frequency response, respectively. We note the relationship $A_{n}e^{j\psi_{n}}=A_{n}e^{j(\phi_{n}+\bar{\psi}_{n})}=e^{j\phi_{n}}h_{n}$.

Assuming the CSI (in the form of the frequency-domain response $h_{n}$ for all $n$) is available to the transmitter, the SMF waveform strategy is a low-complexity method to find the set of amplitudes and phase $\mathbf{s}$, $\Phi$ which maximizes rectenna output power. 
The complex weight on $n^{th}$ SMF sinewave can be given in closed form as
\begin{equation}
 \omega_{n} = e^{-j\bar{\psi}_{n}}A_{n}^{\beta}\sqrt{\frac{2P}{\sum_{n=0}^{N-1}A_{n}^{2\beta}}} .
\end{equation}
The SMF waveform design is only a function of a single parameter, namely $\beta$ [8].

The result of the simulation with the SMF method shows that the performance with $\beta=3$ is close to nearly 95\% of the optimal waveform at -20dBm input power with 8 uniformly spaced sinewaves [8]. The SMF waveform design method will be implemented in our prototype waveform optimizer design in the next section.

\section{Prototyping a Closed-Loop WPT System}
We established a point-to-point wireless power transfer experiment setup with a waveform optimization feature. The system requirement is that the transmitter needs to transmit a pilot and a wireless power signal, and calculate waveform parameters dynamically as a function of the CSI. The receiver needs to estimate the channel and report back to the transmitter, and receive the RF signal and convert it to DC power.

\subsection{System Architecture and equipment setup}
The system operates on the 2.4GHz carrier frequency, which is the same as WiFi in the ISM band. The target distance between the transmitter and receiver is 5m. A transmission signal will be divided into two parts: pilot signal for channel estimation, and channel-optimized power waveform for WPT, respectively. The time period for channel estimation and wireless power transfer can be modified according to the channel coherence time. Currently, the length of one time-slot is set to one second. Because our measurements were carried out in a static office environment, the channel does not change too quickly. Fig.1 indicates the signal structure.
\begin{figure}[!h]
	\centering
	\includegraphics[width=0.4\textwidth]{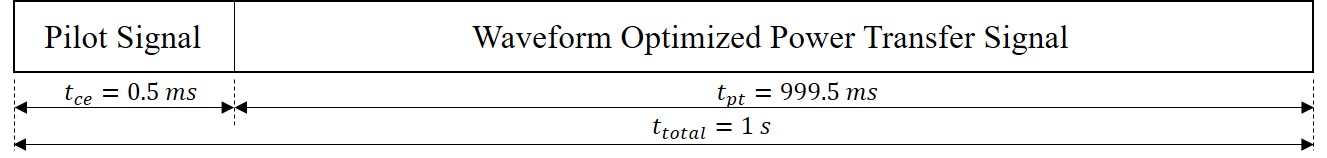}
	\caption{Signal structure of WPT architecture, $t_{ce}$ and $t_{pt}$ denote period of channel estimation and power transfer, respectively.}
	\label{fig1}
\end{figure}

To implement a wireless power and pilot signal transmitter, we choose to use existing Software Defined Radio (SDR) prototyping equipment: National Instrument (NI) FlexRIO (PXI-7966R) and transceiver module (NI 5791R). The NI 5791R operates at 400 MHz to 4.4GHz carrier frequency, and -56 to 8 dBm output power [9]. Its Tx and Rx blocks operate independantly and simultaneously, so we decided to use this SDR also as a channel estimator which receives a pilot signal and estimates the CSI. It simplified our prototyping work, because over-the-air CSI feedback communication is not needed. Therefore, the receiver is divided into two parts - channel estimator (NI equipment) and energy harvester (external rectenna) - and the received RF signal is divided using an external power splitter (ZAPD-2-272+). In addition, because the maximum 8 dBm output power of the equipment cannot meet the wireless power transmission distance of 5m, an external power amplifier (ZHL-16W-43+), which provides 45dB of gain and 41dBm of P1dB, is added in the system [10]. Universal WiFi (2.4GHz band) antennas are used for Tx and Rx, and its gain is 3 dBi. Figs.2 and 3 show system diagrams and installed equipment settings.
\begin{figure}[!h]
	\centering
	\includegraphics[width=0.44\textwidth]{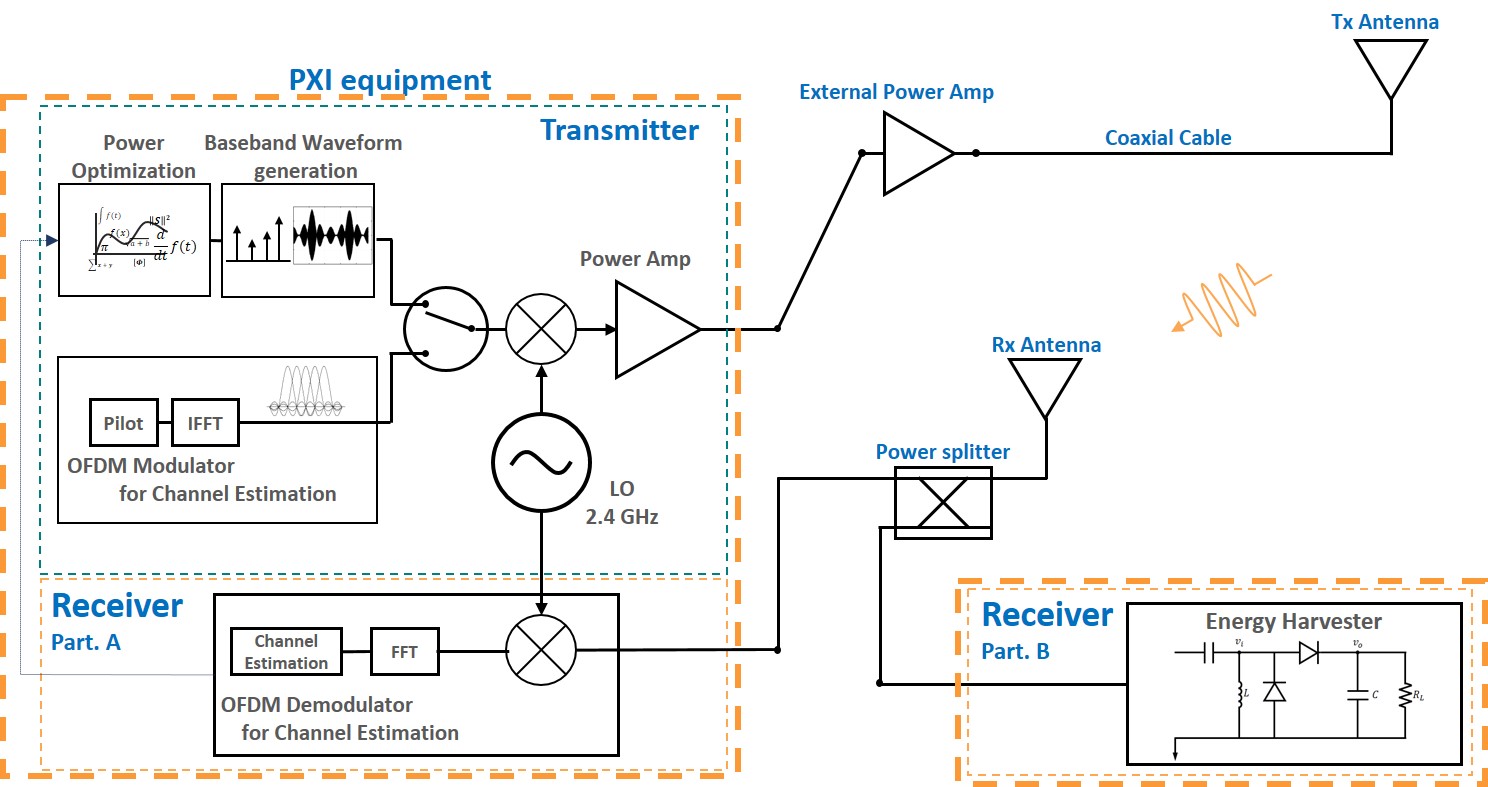}
	\caption{System diagram with equipment and peripherals connection.}
	\label{fig2}
\end{figure}
\begin{figure}[!h]
	\centering
	\includegraphics[width=0.37\textwidth]{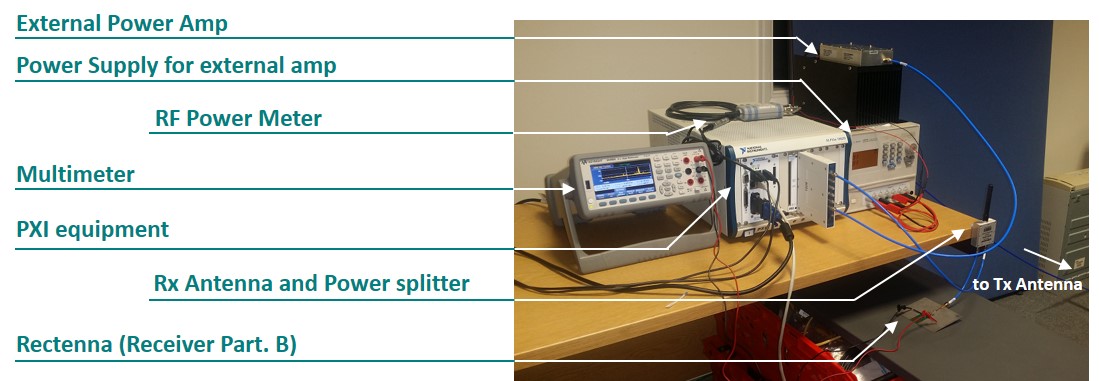}
	\caption{Installed WPT with waveform optimization system.}
	\label{fig3}
\end{figure}

\subsection{Wireless Power Transmitter and Channel Estimator Design}
Channel estimation is a widely used method to restore a transmission signal at the receiver in a communication system. A receiver could recognize the distortion of the transmitted signal during propagation in the channel and restore original signal by channel estimation process. In this study, the CSI will be collected by the same method as in a communication system. However, the CSI is then used to optimize the wireless power transmission waveforms. The pilot-based channel estimation technique used in OFDM systems is implemented in our prototype. OFDM is used to transmit the pilot signal on the same frequencies as the one used by the WPT multisine waveforms. Leveraging channel estimation for OFDM, we can therefore collect CSI on each frequency. The pilot-based technique use a reference pilot signal that the transmitter and the receive both know. The channel status information can be calculated from comparison between received signal and known reference signal at the receiver. We use a block type pilot that allocates pilot symbols on all subcarriers, such that the CSI on all subcarriers can be calculated without interpolation. Also, we use a Least-Square (LS) method to calculate the CSI because of its low complexity. Our channel estimation system operates at 2.4GHz center frequency with a 20MHz bandwidth and 256 subcarriers, corresponding to a subcarrier spacing of 78.125KHz. However, the upper and lower 5MHz bands are used as guard bands, and the region that is actually used to estimate the channel is the 10MHz in the middle. OFDM channel estimation specifications are shown in Table1.

\begin{table}[!h]
\renewcommand{\arraystretch}{1.2}
\caption{Operation parameters of OFDM channel estimation}
\label{table_example}
\centering
\begin{tabular}{c|c}
\hline
Parameter & Value\\
\hline
\hline
Bandwidth & 20 MHz\\
\hline
Number of Subcarriers & 256\\
\hline
Frequency Spacing & 78.125 KHz \\
\hline
Pilot type & Block type pilot \\
\hline
Number of symbols for channel estimation & 20 symbols (320 $\mu s$)\\
\hline
Method of Channel Estimation & Least-square\\
\hline
\end{tabular}
\end{table}

The obtained CSI across the 10MHz bandwidth is passed to a waveform optimizer. Then the waveform optimizer extracts channel information of a desired frequency band and calculates optimal waveform parameters by using the SMF method equation (5). In this study, wireless power transmission signal is designed with 8 uniformly spaced sinewaves in the 10MHz bandwidth. Both channel estimation and waveform optimization blocks are implemented and operated in the NI SDR equipment and the detailed functions are implemented by LabView software. Fig.4 shows sample results of the estimated channel frequency response and 8-multisine WPT waveform amplitudes. Following [6,7,8], more power is allocated to frequency components of the multisine waveform that correspond to large frequency-domain channel gains.
\begin{figure}[!h]
	\centering
	\includegraphics[width=0.36\textwidth]{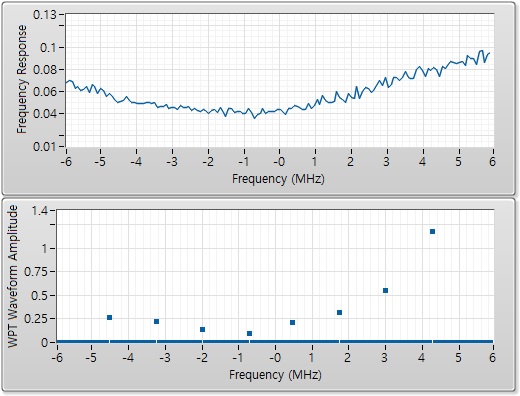}
	\caption{Frequency response of the wireless channel and WPT waveform magnitude (N=8) for 10MHz bandwidth.}
	\label{fig4}
\end{figure}

\subsection{Rectifier Design}
The voltage doubler rectifier circuit is designed by using a Skyworks SMS7630 Schottky diode. The rectifier input is matched to 50 ohms by using T-matching network consisting of series $L_{1}=3.9nH$, shunt $C_{1}=2.2pF$, and series microstrip line width=$0.2mm$ and length=$9.997mm$. Rectifier input $S_{11}$ at 2.4GHz is -30dBm. Load resistance is 5.1 Kohms. The prototyped rectenna RF-to-DC conversion efficiency reaches 12\% at -20dBm continuous wave(1-tone) input. Fig.5 is the circuit schematic and photograph of the fabricated rectenna.
\begin{figure}[!h]
	\centering
	\includegraphics[width=0.43\textwidth]{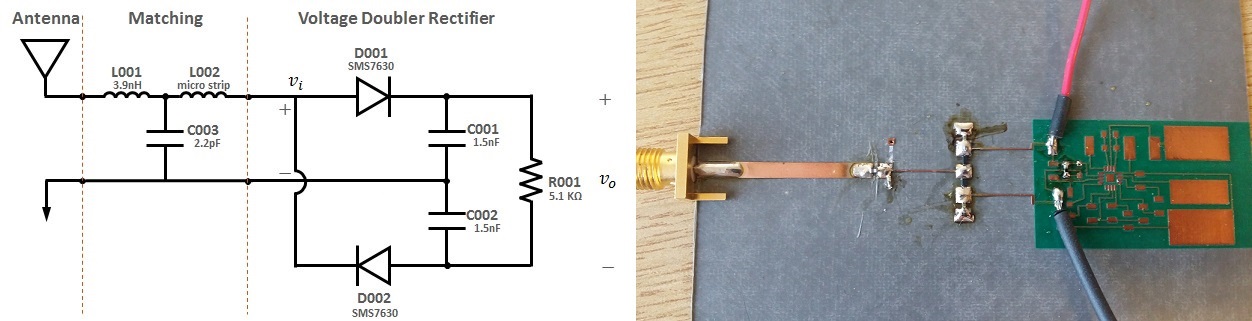}
	\caption{Fabricated rectenna and circuit schematic.}
	\label{fig5}
\end{figure}

\section{Measurement Results}
The performance comparison between channel adaptive and non-adaptive waveforms was conducted in three different cases. Both waveforms are designed with 8 uniformly spaced sinewaves within a 10MHz bandwidth at 2.4GHz center frequency. The channel adaptive waveform parameters are obtained by using SMF method, and the non-adaptive waveform has uniform power allocation and zero phase. Measurements were carried out in a normal office environment in static conditions. In all three cases, transmit power was set to 35dBm and measured RF power at the receiver varied from -19 to -28dBm. We have measured an average harvested DC power for two minutes. Fig.6 indicates measurement locations for three cases. 
\begin{figure}[!h]
	\centering
	\includegraphics[width=0.41\textwidth]{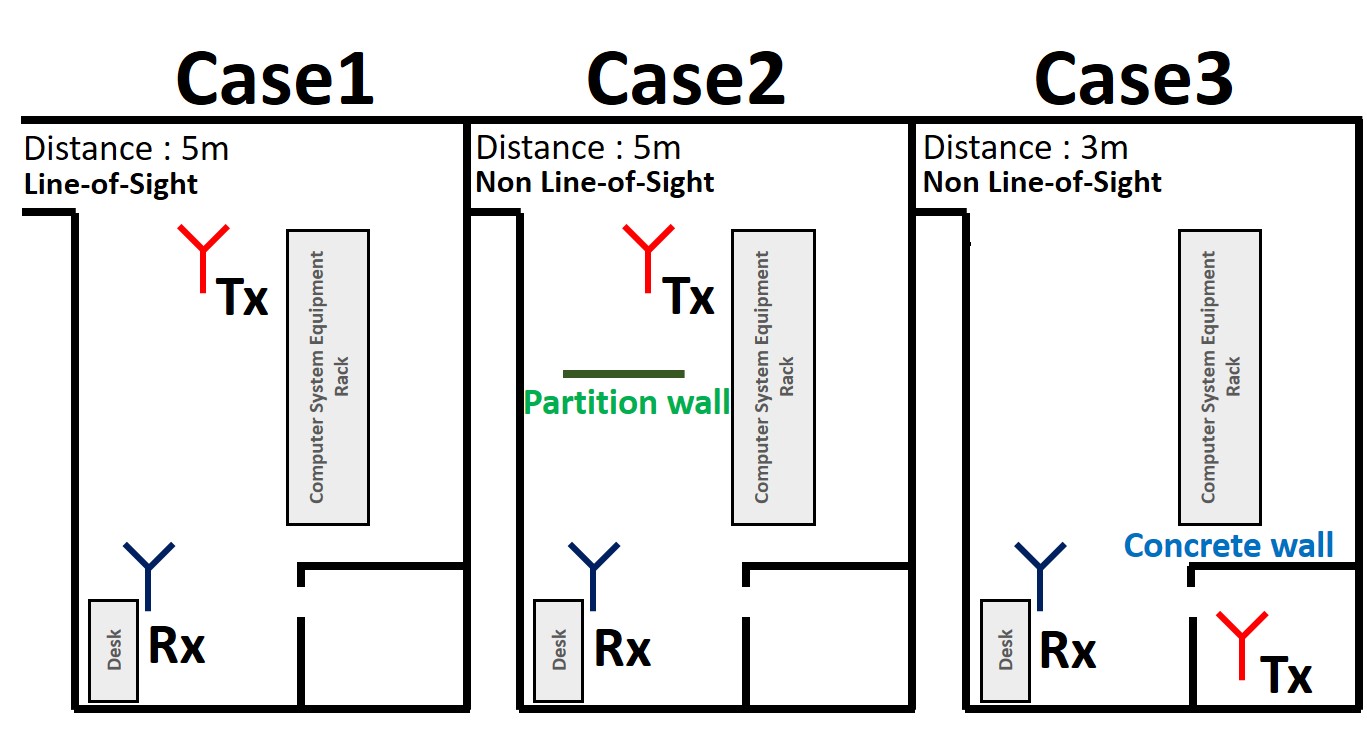}
	\caption{Three different measurements environment}
	\label{fig6}
\end{figure}

The measurements were carried out ten times at the same location while maintaining static conditions and the results are averaged. Experimental results are as shown in Table.2. 

\begin{table}[!h]
\centering
\renewcommand{\arraystretch}{1.2}
\renewcommand{\tabcolsep}{1.5mm}
\renewcommand\thefootnote{\it\alph{footnote}}

\caption{DC power harvesting performance measurement results}
\label{table2}
\begin{minipage}{\textwidth}
\begin{tabular}{c|c|c|c|c|c|c}

\multirow{2}{*}{}	& \multicolumn{2}{c|}{Case 1}	& \multicolumn{2}{c|}{Case 2}	& \multicolumn{2}{c}{Case 3}\\
\cline{2-7} 
                                                                & N\footnote{Non-adapative waveform} & A\footnote{Adaptive Waveform} & N\footnotemark[1] & A\footnotemark[2] & N\footnotemark[1] & A\footnotemark[2] \\

	 \hline
\multicolumn{1}{p{2.9cm}|}{Harvested Power ($\mu$W)} & 1.233                                                   & 1.354    & 0.566                                                   & 0.713    & 1.032                                                   & 1.412    \\ \hline
\multicolumn{1}{p{2.9cm}|}{Performance gain}   & \multicolumn{2}{c|}{9.8\%}                                         & \multicolumn{2}{c|}{25.9\%}                                        & \multicolumn{2}{c}{36.8\%}                                        \\ 
\end{tabular}
\end{minipage}
\end{table}
Experimental results show that the use of channel adaptive waveform shows better results in terms of energy harvesting performance. Performance improvements (in terms of output DC power) of 9.8 to 36.8\% were confirmed through measurements in a static office environment. The result of the test case 1 (Line-of-Sight(LoS)) shows the smallest performance gain. It is because the channel in LoS case is almost frequency flat, and this confirms theoretical observations from [6,7]. On the other hand, when the channel is becoming more frequency-selective (as in NLoS), using adaptive waveform shows much better performance than non-adaptive waveform. So, the performance gain (adaptive vs non-adaptive) in frequency-selective channels is higer than in frequency flat channels.

\section{Conclusion}
A closed-loop and adaptive WPT architecture with channel acquisition and waveform optimization was designed and tested. The DC power harvesting performance was compared between channel adaptive and non-adaptive waveform design. It is shown that using channel adaptive waveform under a fixed transmit power constraint could deliver more power to the receiver, and the gain varies depending on the channel conditions. In accordance with theoretical results from [7], larger gains have been observed in the presence of frequency selective channel as in NLoS conditions. This research is a first step to verify the real-world benefits of using a closed-loop WPT architecture based on channel-adaptive waveforms and channel acquisition at the transmitter. It will be modified to increase the accuracy of channel estimation and reduce the system complexity in order to decrease the processing time, and also expanded to a multi-antenna prototype and a simultaneous wireless information and power transfer (SWIPT) system.






%

\end{document}